\documentclass[twocolumn]{aastex62}
\usepackage{amsmath}

\usepackage{float}
\graphicspath{{./}{figures/}}

\shorttitle{Explaining Neptune's Eccentricity}
\shortauthors{Siraj \& Loeb}

\begin{document}
\title{Explaining Neptune's Eccentricity}

\email{amir.siraj@cfa.harvard.edu, aloeb@cfa.harvard.edu}

\author{Amir Siraj}
\affil{Department of Astronomy, Harvard University, 60 Garden Street, Cambridge, MA 02138, USA}

\author{Abraham Loeb}
\affiliation{Department of Astronomy, Harvard University, 60 Garden Street, Cambridge, MA 02138, USA}




\begin{abstract}

Early migration damped Neptune's eccentricity. Here, we assume that the damped value was much smaller than the value observed today, and show that the closest flyby of $\sim 0.1 \;  \mathrm{M_{\odot}}$ star over $\sim 4.5 \mathrm{\; Gyr}$ in the field, at a distance of $\sim 10^3 \mathrm{\; AU}$ would explain the value of Neptune's eccentricity observed today.

\end{abstract}


\section{Introduction}
Neptune's eccentricity was likely damped within a $\sim 10^8 \mathrm{\; yr}$ timescale \citep{2012ApJ...750...43D, 2014Icar..232...81M, 2015AJ....150...68N, 2018Icar..306..319G, 2019Icar..334...89R, 2021ApJ...908L..47N}, comparable to the lifetime of the Sun's birth cluster \citep{2010ARA&A..48...47A}. Here, we assume that the damped value was much smaller than the value observed today ($e = 9 \times 10^{-3}$) and investigate whether a impulsive interaction with a passing star in the field over the past $\sim 4.5 \mathrm{\; Gyr}$ explains the observed value. 

\section{Methods}
Consider a population of perturbers with mass density $\rho/m \equiv (1/m_p) \; \mathrm{d \ln{\rho_p} / d \ln{m_p}}$, where $\mathrm{d \rho_p / d m_p}$ is the mass function of possible perturbers (stars and lower-mass interstellar objects). individual mass $m$, number density $n \equiv \rho/m$, and velocity dispersion $\sigma$. We adopt the impulse approximation when the interaction timescale is shorter than the orbital period. In the regime where the impact parameter $b$ exceeds Neptune's semi-major axis $a$, namely $b > b_{min} = a$, this condition restricts the impact parameter to,
\begin{equation}
b < b_{max} = 2 \pi \sigma \sqrt{\frac{a^3}{G M_{\odot}}} \; \; .
\end{equation}

The limit on perturber number density $n$ is then set by requiring at least one interaction over timescale $\tau$ within the impact parameter $b_{max}$, 
\begin{align}
\begin{split}
    n \; \pi \; b_{max}^2 \; \sigma \; \tau > 1 , 
\end{split}
\end{align}
which can be expressed as,
\begin{align}
\begin{split}
    \label{eq:max}
n & > \;  (\pi  \; \sigma \; \tau)^{-1} \; \times (2 \pi \sigma)^{-2} (G M_{\odot} / a^3) \\ & = \; G M_{\odot} \; / \; 4 \pi^3 \; \sigma^3 \; a^3 \;  \tau \; \; \\ & = \; 0.09 \; \mathrm{pc^{-3}} \; \left( \frac{\sigma}{30 \; \mathrm{km \; s^{-1}}}\right)^{-3} \; \\ & \; \; \; \; \left(\frac{a}{30 \; \mathrm{AU}}\right)^{-3} \; \left( \frac{\tau}{4.5 \; \mathrm{Gyr}}\right) \; \; .
\end{split}
\end{align}
We adopt a fiducial value of $\sigma \sim 30 \mathrm{\; km \; s^{-1}}$ since it characterizes the three-dimensional velocity dispersion of local Galactic disk stars relative to the Sun. We derived this value through a Monte Carlo simulation, given the Sun's velocity vector relative to local standard of rest (LSR) of, $(v_U^{\odot}, v_V^{\odot}, v_W^{\odot}) = (10 \pm 1, 11 \pm 2, 7 \pm 0.5) \; \mathrm{km \; s^{-1}}$ \citep{2010MNRAS.403.1829S, 2015ApJ...809..145T, 2016ARA&A..54..529B} and the velocity dispersion of local stars around the LSR of, $(\sigma_U, \sigma_V, \sigma_W) = (33 \pm 4, 38 \pm 4, 23 \pm 2) \; \mathrm{km \; s^{-1}}$ \citep{2018MNRAS.474..854A}.

Requiring $b_{max} > b_{min}$ gives an upper limit on the perturber number density, 
\begin{align}
\begin{split}
    \label{eq:min}
n & < (\pi  \; \sigma \; a^2  \; \tau)^{-1} \; \; \\ & = \; 109 \; \mathrm{pc^{-3}} \; \left( \frac{\sigma}{30 \; \mathrm{km \; s^{-1}}}\right)^{-3} \; \\ & \; \; \; \; \left(\frac{a}{30 \; \mathrm{AU}}\right)^{-2} \; \left( \frac{\tau}{4.5 \; \mathrm{Gyr}}\right)^{-1} \; \; .
\end{split}
\end{align}

In the regime considered here, namely $b > a$, the typical impulsive interaction between a perturber and Neptune results in a velocity kick to the planet, $\Delta v_{N} \approx (4 \; G \; m \; a \; / \; v \; b^2) $ \citep{2009ApJ...697..458S}. Starting from $e \ll \Delta e$, a single encounter delivers a change to Neptune's orbital eccentricity $e$ as follows \citep{2009ApJ...697..458S},
\begin{equation}
\begin{split}
    (\Delta e)^2   & \sim \; \Delta v_{N} / v_{N} \\ & = 4 \; \sqrt{\frac{G a^3}{M_{\odot} \sigma^2}} \left( \frac{m}{b^2} \right) \; \; ,
\end{split}
\end{equation}
where $v_N$ is Neptune's orbital speed. Since the closest encounter requires,
\begin{equation}
\begin{split}
    \frac{m}{b^2} = \pi \rho \sigma \tau \; \; ,
\end{split}
\end{equation}
the change to orbital eccentricity can be expressed as 
\begin{equation}
\begin{split}
    \label{eq:ede2}
    (\Delta e)^2 \sim \;  4 \pi \rho \tau \; \sqrt{\frac{G a^3}{M_{\odot}}} \; .
\end{split}
\end{equation}
\section{Results}
To produce Neptune's observed eccentricity $e$, the mass density $\rho$ of the perturber population must follow,
\begin{equation}
\begin{split}
    \label{eq:rho}
    \rho \sim \frac{(\Delta e)^2}{4 \pi \tau} \; \sqrt{\frac{M_{\odot}}{G a^3}} \; \; ,
\end{split}
\end{equation}
or,
\begin{equation}
\begin{split}
    \label{eq:rho_n}
    \rho \sim \; & 1.2 \times 10^{-2} \; \mathrm{M_{\odot} \; pc^{-3}}\;\left( \frac{\Delta e}{9 \times 10^{-3}}\right)^2 \\ & \left( \frac{\tau}{4.5 \; \mathrm{Gyr}}\right)^{-1} \left( \frac{a}{30 \; \mathrm{AU}}\right)^{-3/2}  \; .
\end{split}
\end{equation}

\begin{figure}
  \centering
  \includegraphics[width=\linewidth]{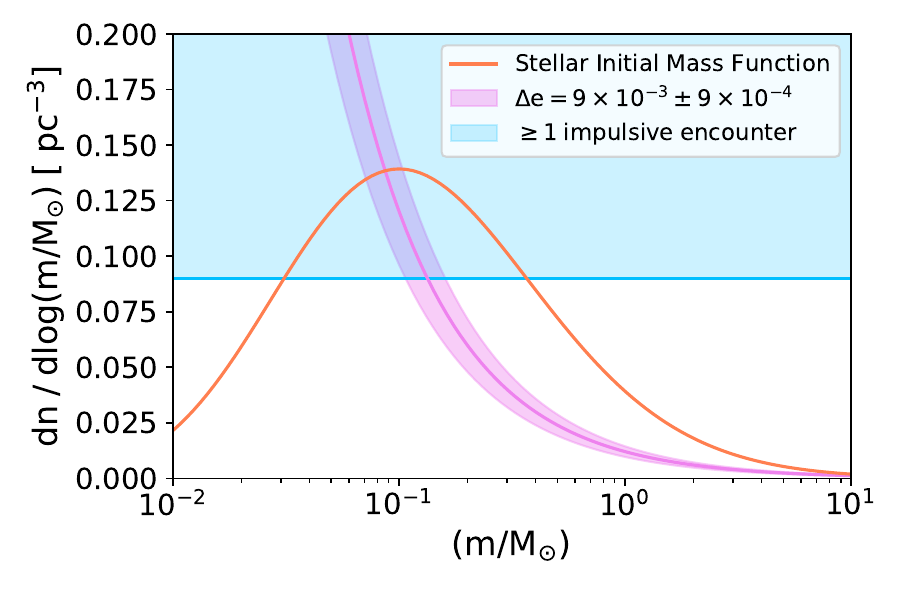}
    \caption{A star with mass $0.09 \pm 0.02 \; \mathrm{M_{\odot}}$ was capable of producing Neptune's eccentricity, since the range lies at the intersection of the IMF (orange, \citep{2003PASP..115..763C}) with the $\Delta e$ condition (pink, Eq. \ref{eq:rho_n}) and the requirement of having at least one impulsive encounter over $\tau \sim 4.5 \; \mathrm{Gyr}$ (blue, Eq. \ref{eq:max}).}
    \label{fig:ups}
\end{figure}

Figure \ref{fig:ups} illustrates the parameter space over which unbound planets are excluded by the observed eccentricity of Neptune. The closest encounter of a star with mass $0.09 \pm 0.02 \; \mathrm{M_{\odot}}$ over $\tau \sim 4.5 \mathrm{\; Gyr}$ in the field would explain Neptune's eccentricity today, assuming it was initially damped to some smaller value.

We only consider the single closest encounter here since at the peak of the distribution ($n \sim 0.1 \mathrm{\; pc^{-3}}$) the minimum impact parameter is comparable to $b_{max} \sim 10^3 \mathrm {\; AU}$ (see Eq. \ref{eq:max}). Encounters at larger distances would make negligible contributions to Neptune's eccentricity since they would lie in the adiabatic regime. If any other impulsive encounters were to take place, even though statistically unlikely, they would add in quadrature due to the resulting random walk in eccentricity, changing the result only by order unity. Furthermore, we only consider dwarf stars because they exclusively satisfy the condition on stellar number density in Equation \eqref{eq:max}.

\vspace{0.2in}
We thank S. Tremaine, J. Miralda-Escudé, A. Riotto, G. Franciolini, and V. De Luca for insightful comments on an early version of a related manuscript. This work was supported in part by the Black Hole Initiative at Harvard University, which is funded by JTF and GBMF. 






\bibliography{bib}{}
\bibliographystyle{aasjournal}



\end{document}